\documentclass[epj,nopacs]{svjour}
\usepackage{graphics}
\usepackage{threeparttable}
\usepackage{longtable,booktabs}
\usepackage{multirow}
\begin{document}
\title{Correlation, Breit and Quantum Electrodynamics effects on energy level and transition properties of W$^{54+}$ ion}
\author{XiaoBin Ding\inst{1,}\thanks{Email:dingxb@nwnu.edu.cn} \and Rui Sun\inst{1} \and Fumihiro Koike\inst{2} \and Daiji Kato\inst{3} \and Izumi Murakami\inst{3} \and Hiroyuki A Sakaue\inst{3} \and Chenzhong Dong\inst{1}
}                     
%
%
\institute{Key Laboratory of Atomic and Molecular Physics and Functional Materials
of Gansu Province, College of Physics and Electronic Engineering, Northwest Normal University, Lanzhou 730070, China.   \and Department of physics, Sophia University, Tokyo 102-8554, Japan \and National Institute for Fusion Science, Toki, Gifu 509-5292, Japan}
%
\date{Received: date / Revised version: date}
%
\abstract{The electron correlation effects and Breit interaction as well as Quantum Electro-Dynamics (QED) effects were expected to have important contribution to the energy level and transition properties of heavy highly charged ions. The ground states [Ne]$3s^{2}3p^{6}3d^{2}$ and first excited states [Ne]3s$^{2}3p^{5}3d^{3}$ of W$^{54+}$ ion have been studied by using Multi-Configuration Dirac-Fock method with the implementation of Grasp2K package. A restricted active space was employed to investigate the correlation contribution from different models. The Breit interaction and QED effects were taken into account in the relativistic configuration interaction calculation with the converged wavefunction. It is found that the correlation contribution from 3s and 3p orbital have important contribution to the energy level, transition wavelength and probability of the ground and the first excited state of W$^{54+}$ ion.
%
} 

\titlerunning{Electron correlation, Breit, and QED effects on W$^{54+}$ ions }
\authorrunning{Ding XB, Sun R et.al.}
\maketitle{}
\section{Introduction}
\label{intro}
Tungsten was chosen to be used as the armour material of the divertor of the ITER project due to its favourable physical and chemical properties. The generation of the tungsten ions in various different ionization stages during the plasma-wall interaction and plasma transport will be an inevitable issue. Even in the ITER core plasma with high temperature ($T_e \sim 25 keV$), the tungsten ions could not be fully ionized. Therefore, large radiation loss from the highly charged tungsten ions could be expected which will lead to the disruption of the plasma if the relative concentration of W ion impurities in the core plasma is higher than about $10^{-5}$ \cite{11}. There is thus a strong demand for energy levels and transition properties of tungsten ions.

The study of W$^{54+}$ ion not only provide necessary reference data for the fusion plasma physics but also provide a chance to study the fundamental physics. The ground states [Ne]$3s^{2}3p^{6}3d^{2}$ and the first excited states [Ne]3s$^{2}3p^{5}3d^{3}$ of W$^{54+}$ ion are highly charged multi-electron ions. Therefore, the relativistic effects, electron correlation effects as well as the Breirt and Quantum Electro-dynamic (QED) effects should be taken into account to calculate the energy level structure and transition properties.

In recent several decades, several theoretical and experimental work have been performed on the energy level and  the E1, E2, M1 transitions properties of W$^{54+}$ ion \cite{PhysRevA.63.032518,0953-4075-41-2-021003,0953-4075-43-7-074026,Ralchenko2011,0953-4075-44-19-195007,0953-4075-48-14-144020,Lennartsson2013,doi:10.1139/cjp-2014-0636}. U. I. Safronova et al. calculated 9 fine structure levels belonging to the ground state configuration [Ne]3s$^{2}$3p$^{6}$3d$^{2}$ of W$^{54}$ ions, and the M1 and E2 transitions between these levels also have been studied by using the relativistic many-body perturbation theory (RMBPT) \cite{0953-4075-43-7-074026}. They started their calculations from 1s$^{2}$2s$^{2}$2p$^{6}$3s$^{2}$3p$^{6}$ Dirac-Fock potential. P. Quinet calculated the energy levels and the forbidden transition of the ground configuration for several highly charged tungsten ions ($W^{47+}-W^{61+}$) by using multi-configuration Dirac-Fock (MCDF) method. The correlation within the n = 3 complex and some $n=3 \rightarrow n'=4$ single excitations were considered in their calculations \cite{0953-4075-44-19-195007}. Furthermore, X. L. Guo et al calculated the energy, transition wavelength and probability for forbidden transition within the $3d^k$ ground configuration of highly charged Hf, Ta, Au and W ions by RMBPT and the relativistic configuration-interaction (RCI) method \cite{0953-4075-48-14-144020}. Y. Ralchenko et al. observed the M1 spectrum from 3d$^{n}$($n$$=$1-9) ground state fine structure multiplets of tungsten ions with electron-beam ion trap (EBIT) and a non-Maxwellian collisional-radiative model was used to analyze the observed spectrum \cite{Ralchenko2011}. The configuration interaction among n=3 complex and the single excitation up to n=5 was included in their calculation. This experiment was helpful to reveal the energy levels of these ions. T. Lennartsson et al. observed the E1 spectrum of the excited state [Ne]3s$^{2}$3p$^{5}$3d$^{3}$ to the ground state [Ne]3s$^{2}$3p$^{6}$3d$^{2}$ of W$^{54+}$ ion in the wavelength range of 26.5-43.5{\AA} by using EBIT \cite{Lennartsson2013}. A collisional-radiative model was applied to explain the observed spectrum. Finally, an MCDF calculation with restricted electron correlation effects on E1 transitions was presented by Dipti et al \cite{doi:10.1139/cjp-2014-0636}.

Most of the previous studies are based on the relativistic theory which was necessary to investigate the heavy highly charged ions. However, only limited electron correlation effects were taken into account in the previous calculation. The electron correlation, Breit interaction and QED effects on the energy level, transition wavelength and probability need to be investigated further.

\section{Theoretical method and computational methodology}
\label{sec:1}
 Multi-configuration Dirac-Fock method which based on the fully relativistic theory had been successfully used to investigate the complex atoms\cite{0953-4075-44-14-145004,Ding2016,Aggarwal2016187,Ding2016874,Aggarwal2016205,Aggarwal20159,Ding2012,Chen2017258,Singh2016339,Wang201615}. The details of the MCDF theory was expounded by I.P. Grant \cite{Grant2007}. The GRASP family code which based on MCDF theory had been developed by a couple of authors \cite{Grant1980,Desclaux1984,K.1989,GRASP92,51934790}. The present calculation was done by using the newly developed version of GRASP2K by P. J\"{o}nssons \cite{51934790}. The investigations were performed by a systematic restricted active space (RAS) method to study the electron correlation effects.

In RAS, all the electron of the interested atom can be divided into 3 types, valence, core and inactive core electron. The correlation effects among them will contribute to the total energy of the atoms. However, for the fine energy structure and transition probability the correlation contributions from inactive core were expected to be small due the cancellation effects. Take the ground states [Ne]$3s^{2}3p^{6}3d^{2}$ as an example, the 'Ne' core, i.e., $1s^2 2s^2 2p^6$ was regarded as inactive core which means that no electron in the inactive core will be excited to the virtual orbital to construct the correlation configuration state functions (CSFs), while $3d$ orbital was regarded as valence electron in all the calculation. For the core electron, $3p$ and $3s 3p$ are regarded as core electron respectively which were called \emph{Model 1} and \emph{2}, to investigate the correlation contribution from differen core subshells.

The calculation was started from a single configuration Dirac-Fock solution. Then the computation was performed by increase the virtual orbital layer-by-layer systematically. Only the new added orbital will be optimized each time. With the increase of the virtual orbital, the number of CSF increased rapidly. In order to keep the calculation manageable and traceable, the virtual orbital extended to the principal quantum number n up to 5.

After the converged energy level and wavefunction has been obtained, a configuration interaction (CI) calculation had been performed with Breit interaction and QED effects added. And then the E1 transition properties were calculated by using wavefunction obtained by biorthogonal transformation technique to include the relaxation effects which due to the sudden change of the potential for the initial and final state of the transition.

\section{Results and discussion}
\label{sec:2}
\subsection{The ground configuration 3p$^{6}$3d$^{2}$ of W$^{54+}$ ion}\label{sec:2.1}

\begin{table}
\renewcommand\arraystretch{1.1}
\renewcommand\tabcolsep{4pt}
\footnotesize
\caption{The 9 energy levels (in eV) of the ground configuration [Ne]3s$^{2}$3p$^{6}$3d$^{2}$ of W$^{54+}$ ion calculated from different correlation models. The notation \lq\lq Index\rq\rq \ is the level index sorted by energy which could be found in the Table 1 of \cite{DingXB2016}. \lq \lq DF \rq \rq is the single configuration Dirac-Hartree-Fock calculation, while other columns were the results from different active space, see the details in Sec.\ref{sec:2.1}.} \label{Tab1}

\begin{threeparttable}
\begin{tabular}{c c c c c c}

\toprule
 	&	&	&	Energy(eV)	&	&		\\
\cline{1-6}

 Index	& DF	&	n=3	&	n=4	&	n=5	&	n=5(5s-5d)	\\
\hline
\multicolumn{6}{c}{Model 1 (VV)}\\
	1	&	0.000 	&	0.000 	&	0.000 	&	0.000 	&	0.000 	\\

    2	&	23.467 	&	23.464 	&	23.156 	&	23.070 	&	23.122 	\\
	3	&	73.876 	&	73.876 	&	73.917 	&	73.939 	&	73.921 	\\
	4	&	84.790 	&	84.788 	&	84.698 	&	84.680 	&	84.695 	\\
	5	&	88.844 	&	88.842 	&	88.843 	&	88.797 	&	88.835 	\\
	6	&	89.048 	&	89.045 	&	88.915 	&	88.912 	&	88.915 	\\
	7	&	156.490 	&	156.490 	&	156.511 	&	156.504 	&	156.510 	\\
	8	&	164.576 	&	164.574 	&	164.459 	&	164.426 	&	164.454 	\\
	9	&	188.879 	&	188.871 	&	188.338 	&	188.154 	&	188.271 	\\
\cline{1-6}
\multicolumn{6}{c}{Model 1 (VV+CV)}\\
	1	&	0.000 	&	0.000 	&	0.000 	&	0.000 	&	0.000 	\\
2	&	23.467 	&	23.464 	&	23.510 	&	23.422 	&	23.475 	\\
3	&	73.876 	&	73.876 	&	73.834 	&	73.852 	&	73.840 	\\
4	&	84.790 	&	84.788 	&	84.732 	&	84.709 	&	84.731 	\\
5	&	88.844 	&	88.842 	&	88.656 	&	88.618 	&	88.663 	\\
6	&	89.048 	&	89.045 	&	88.934 	&	88.922 	&	88.931 	\\
7	&	156.490 	&	156.490 	&	156.354 	&	156.346 	&	156.366 	\\
8	&	164.576 	&	164.574 	&	164.489 	&	164.448 	&	164.492 	\\
9	&	188.879 	&	188.871 	&	188.988 	&	188.795 	&	188.938 	\\

\cline{1-6}
\multicolumn{6}{c}{Model 1 (VV+CV+CC)}\\
	1	&	0.000 	&	0.000 	&	0.000 	&	0.000 	&	0.000 	\\
	2	&	23.467 	&	22.570 	&	22.593 	&	22.504 	&	22.559 	\\
	3	&	73.876 	&	74.177 	&	74.126 	&	74.144 	&	74.130 	\\
	4	&	84.790 	&	84.738 	&	84.652 	&	84.628 	&	84.649 	\\
	5	&	88.844 	&	88.939 	&	88.782 	&	88.765 	&	88.775 	\\
	6	&	89.048 	&	89.172 	&	88.940 	&	88.898 	&	88.943 	\\
	7	&	156.490 	&	156.961 	&	156.788 	&	156.777 	&	156.794 	\\
	8	&	164.576 	&	164.537 	&	164.412 	&	164.369 	&	164.411 	\\
	9	&	188.879 	&	187.247 	&	187.322 	&	187.133 	&	187.271 	\\
\cline{1-6}
\multicolumn{6}{c}{Model 2 (VV+CV)}\\
	1	&	0.000 	&	0.000 	&	0.000 	&	0.000 	&	0.000 	\\
  2	&	23.467 	&	23.296 	&	23.313 	&	23.276 	&	23.278 	\\
3	&	73.876 	&	73.947 	&	73.911 	&	73.918 	&	73.918 	\\
4	&	84.790 	&	84.532 	&	84.460 	&	84.453 	&	84.459 	\\
5	&	88.844 	&	88.660 	&	88.519 	&	88.527 	&	88.514 	\\
6	&	89.048 	&	88.831 	&	88.641 	&	88.592 	&	88.651 	\\
7	&	156.490 	&	156.502 	&	156.370 	&	156.349 	&	156.383 	\\
8	&	164.576 	&	164.262 	&	164.161 	&	164.141 	&	164.166 	\\
9	&	188.879 	&	188.896 	&	188.997 	&	188.895 	&	188.951 	\\

\cline{1-6}
\multicolumn{6}{c}{Model 2 (VV+CV+CC)}\\
	1	&	0.000 	&	0.000 	&	0.000 	&	0.000 	&	0.000 	\\
	2	&	23.467 	&	22.341 	&	22.315 	&	22.281 	&	22.278 	\\
	3	&	73.876 	&	74.239 	&	74.197 	&	74.201 	&	74.203 	\\
	4	&	84.790 	&	84.503 	&	84.402 	&	84.399 	&	84.393 	\\
	5	&	88.844 	&	88.583 	&	88.395 	&	88.387 	&	88.398 	\\
	6	&	89.048 	&	89.157 	&	88.924 	&	88.931 	&	88.871 	\\
	7	&	156.490 	&	156.964 	&	156.800 	&	156.809 	&	156.776 	\\
	8	&	164.576 	&	164.248 	&	164.109 	&	164.110 	&	164.087 	\\
	9	&	188.879 	&	187.095 	&	187.109 	&	187.064 	&	187.015 	\\
\bottomrule
\end{tabular}
\begin{tablenotes}
\item[]
\end{tablenotes}
\end{threeparttable}
\end{table}

The effects of electron correlation on the 9 energy levels(eV) of the ground configuration 3p$^{6}$3d$^{2}$ of W$^{54+}$ ion were presented in the Table~\ref{Tab1}. The electron correlation effects were taken into account by dividing the reference configuration [Ne]3s$^{2}$3p$^{6}$3d$^{2}$ into inactive core, core and valence shells. In this way, two models are employed in the present calculations in order to study the importance of electron correlation effects referring to 3s and 3p orbitals. In model 1, 3p orbital was regarded as core, while 3s and 3p orbitals were regarded as core in model 2. Then the valence-valence (VV), core-valence (CV) and core-core (CC) correlations are considered by single and double substituting the occupied electrons in specific orbits to an active orbit sets, which are marked with n=3, 4, 5, 5(5s-5d),respectively. DF in Table ~\ref{Tab1} refers to Dirac-Fock calculation, while n=5(5s-5d) means the active orbital restricted to include {5s, 5p, 5d} only. The configuration spaces were extended layer by layer in the present calculations to evaluate the electronic correlation effects efficiently and circumvent the convergence problem that one frequently encounters in SCF calculations. The Breit interaction and quantum electrodynamics effects (QED) were not included in the calculations in the Table~\ref{Tab1}. In any of the correlation model, the convergence trend can be observed with the increase of the active space. It was shown in the Table \ref{Tab1} that the VV correlation gives the most important contribution to all the energy level. However, in order to obtain more accurate results the correlation effects from other inner orbital also need to be considered. It was found from the Table that the VV+CV contribution from the 3p and {3s,3p} orbital have similar magnitude. This mainly due to the wavefunction expansion from  the restricted active space for VV and VV+CV from 3s and {3s,3p} is almost the same by the restriction of parity and symmetry. However, the VV+CV+CC contribution from these two models are quite different. The energy level obtained by Model 2 obviously lower than that from Model 1. And a better convergence trend could be obtained from the Model 2. Meanwhile, the result from n=5(5s-5d) with VV+CV+CC correlation just have small difference with the one obtained by n=5 correlation model, while the number of CSFs were dramatically  reduced. Therefore, n=5(5s-5d) active space were taken as an economic correlation model for the present calculation.


\begin{table*}
\renewcommand\arraystretch{1.1}
\footnotesize
\centering
\caption{The energy levels (in eV) of the ground configuration [Ne]3s$^{2}$3p$^{6}$3d$^{2}$ of W$^{54+}$ ion with the electron correlation effects, transverse Breit interaction (B) and quantum electrodynamics effects (QED) were taken into account. The notation \lq\lq Index\rq\rq \ is the level index sorted by energy which could be found in the Table 1 of \cite{DingXB2016}. The \lq\lq Coulomb\rq\rq \ represent the results with electron correlation effects only, while the \lq\lq C+B+Q\rq\rq \ represent the results with both electron correlation and the Breit interaction as well as QED effects taken into account. All the calculations were performed in Model 2 with VV+CV+CC correlations. Other available theoretical and experimental results were also listed for comparison. }\label{Tab2}
\begin{threeparttable}
\begin{tabular}{c | c c c | c c c c c c}
\hline

\multirow{2}*{Index}	&		&	Energy(eV)	&		&		&		&	Other(eV)	&		&		\\

\cline{2-9}
	&	Coulomb	&	B+Q (Contri.)	&	C+B+Q	&	NIST$^{a}$	&	MCDF$^{b}$	&	RMBPT$^{c}$	&	RMBPT$^{d}$	&	RCI$^{e}$\\
\hline
1	&	0.000 	&	0.000 	&	0.000 	&	0.000 	&	0.00 	&	0.000 	&	0.000 	&	0.000 	\\

2	&	22.278 	&	0.844 	&	23.123 	&	23.3 	&	23.173 	&	23.199 	&	22.920 	&	24.322 	\\

3	&	74.203 	&	-1.747 	&	72.456 	&	72.590 	&	72.345 	&	72.264 	&	72.630 	&	72.009 	\\

4	&	84.393 	&	-1.588 	&	82.805 	&	82.882 	&	82.772 	&	82.600 	&	82.822 	&	83.076 	\\

5	&	88.398 	&	-2.040 	&	86.358 	&	86.4 	&	86.429 	&	86.021 	&	86.385 	&	86.128 	\\

6	&	88.871 	&	-1.259 	&	87.613 	&	87.9 	&	87.669 	&	87.460 	&	87.629 	&	88.160 	\\

7	&	156.776 	&	-3.619 	&	153.158 	&	153.0 	&	153.009 	&	152.704 	&	153.369 	&	152.591 	\\
8	&	164.087 	&	-2.928 	&	161.160 	&	161.1 	&	161.006 	&	160.774 	&	161.214 	&	161.397 	\\
9	&	187.015 	&	-1.877 	&	185.139 	&	185.1 	&	184.860 	&	184.927 	&	184.883 	&	186.772 	\\

\hline

\end{tabular}
\begin{tablenotes}
\item[] $^a$ From Y. Ralchenko et al by an electron-beam ion trap (EBIT) and an non-Maxwellian collisional-radiative model \cite{Ralchenko2011}
\item[] $^b$ From P. Quinet by MCDF method \cite{0953-4075-44-19-195007}
\item[] $^c$ From U. I. Safronova and A. S. Safronova by RMBPT method \cite{0953-4075-43-7-074026}
\item[] $^d$ From X. L. Guo et al by RMBPT method \cite{0953-4075-48-14-144020}
\item[] $^e$ From X. L. Guo et al by RCI method \cite{0953-4075-48-14-144020}
\end{tablenotes}
\end{threeparttable}
\end{table*}

Generally, Breit interaction and QED effects are expected to be strong in the heavy highly charged ions. The 9 energy levels (in eV) including correlation, Breit interaction and QED effects (C+B+Q) of the ground configuration 3p$^{6}$3d$^{2}$ of W$^{54+}$ ion were presented in Table ~\ref{Tab2} to investigate the contribution from the Breit and QED effects. The other available theoretical and experimental results were also listed for comparison \cite{0953-4075-43-7-074026,Ralchenko2011,0953-4075-44-19-195007,0953-4075-48-14-144020}. The notation \lq\lq C+B+Q\rq\rq \ and \lq\lq Coulomb\rq\rq \ in the 4$^{th}$ and 2$^{th}$ columns of the table represent the results with and without Breit and QED effects which calculated in n=5(5s-5d) model and regarding {3s,3p} orbital as core and VV+CV+CC correlations were taken into account. The notation \lq\lq B+Q (contri.) \rq\rq \ represents the contributions from transverse Breit interaction in low frequency limit and QED effect to the level energies. It can be found in Table ~\ref{Tab2} that the contribution of B+Q is about 2\% for these energy levels. It plays an important role in the calculation of the energy levels of W$^{54+}$ ion. The present calculated energy levels made good agreement with the other available theoretical and experimental results. The minimum relative deviation between the present results and the results from NIST was up to 0.02\%. The satisfactory agreement shows that the model employed in the present work was reliable and the correlation effects have been considered sufficiently. Thus the first excited state 3s$^{2}$3p$^{5}$3d$^{3}$ of W$^{54+}$ ions could also be calculated with the same correlation model of the ground state.

\subsection{The E1 transition properties of W$^{54+}$ ion} \label{sec:1.2}

\begin{table*}
\renewcommand\arraystretch{1.1}
\footnotesize
\centering
\caption{The transition wavelength $\lambda$ (in nm) of some selected E1 transitions between [Ne]3s$^{2}$3p$^{5}$3d$^{3}$ and [Ne]3s$^{2}$3p$^{6}$3d$^{2}$ configurations in W$^{54+}$ ion. The i and j in \lq \lq Index \rq\rq is the energy level index which could be found in the Table 1 of \cite{DingXB2016}. The \lq\lq Coulomb\rq\rq \ represent the wavelength with the electron correlation effects only, while \lq \lq C+B+Q\rq \rq represent the wavelength with electron correlation, Breit and QED effects. Other available experimental and theoretical results were listed for comparison.}\label{Tab3}
\begin{threeparttable}
\begin{tabular}{l c | c c c c | c | c | c c}
\hline

 \multicolumn{2}{c|}{\multirow{2}*{Index}}	&		 \multicolumn{4}{|c|}{Coulomb} 		&	\multirow{3}*{B+Q (Contri.)}	&	C+B+Q	&	 \multicolumn{2}{c}{\multirow{2}*{$\lambda_{Other}$(nm)}}		\\
\cline{3-6}  \cline{8-8}
 &		&		 \multicolumn{4}{|c|}{$\lambda$(nm)}		&		&	$\lambda$(nm)	&		&		\\
\cline{1-6}  \cline{8-10}
i	&	j	&	DF	&	n=3	&	n=4	&	n=5(5s-5d)	&		&	n=5(5s-5d)	&	Exp.$^{a}$	&	Tho.	\\
\hline
& & & & & & & & & \\
1	&	30	&	3.2186 	&	3.2053 	&	3.2100 	&	3.2107 	&	0.0294 	&	3.2401 	&	3.2264 	&	3.2416$^{b}$	\\
	&		&		&		&		&		&		&		&		&	3.2502$^{c}$	\\
1	&	31	&	3.1467 	&	3.1436 	&	3.1490 	&	3.1497 	&	0.0290 	&	3.1787 	&	3.1811 	&	3.1786$^{b}$	\\
	&		&		&		&		&		&		&		&		&	3.1783$^{c}$	\\
1	&	32	&	3.1439 	&	3.1332 	&	3.1399 	&	3.1406 	&	0.0326 	&	3.1732 	&	3.1776 	&	3.1711$^{b}$	\\
	&		&		&		&		&		&		&		&		&	3.1765$^{c}$	\\
1	&	33	&	3.1167 	&	3.1135 	&	3.1219 	&	3.1226 	&	0.0310 	&	3.1536 	&	3.1563 	&	3.1505$^{b}$	\\
	&		&		&		&		&		&		&		&		&	3.1503$^{c}$	\\
1	&	34	&	3.1014 	&	3.0996 	&	3.1084 	&	3.1091 	&	0.0319 	&	3.1410 	&	3.1430 	&	3.1386$^{b}$	\\
	&		&		&		&		&		&		&		&		&	3.1378$^{c}$	\\
2	&	38	&	3.0892 	&	3.0768 	&	3.0907 	&	3.0916 	&	0.0335 	&	3.1251 	&	3.1245 	&	3.1155$^{b}$	\\
	&		&		&		&		&		&		&		&		&	3.1263$^{c}$	\\
1	&	38	&	2.9186 	&	2.9152 	&	2.9278 	&	2.9289 	&	0.0241 	&	2.9530 	&	2.9560 	&	2.9452$^{b}$	\\
	&		&		&		&		&		&		&		&		&	2.9456$^{c}$	\\
6	&	68	&	3.0792 	&	3.0680 	&	3.0809 	&	3.0818 	&	0.0297 	&	3.1115 	&		&		\\
	&		&		&		&		&		&		&		&		&		\\
7	&	79	&	3.0663 	&	3.0595 	&	3.0687 	&	3.0696 	&	0.0251 	&	3.0947 	&		&		\\
	&		&		&		&		&		&		&		&		&		\\
5	&	69	&	3.0651 	&	3.0535 	&	3.0645 	&	3.0654 	&	0.0244 	&	3.0898 	&		&		\\
	&		&		&		&		&		&		&		&		&		\\
8	&	110	&	1.9066 	&	1.9090 	&	1.9110 	&	1.9112 	&	0.0154 	&	1.9266 	&		&		\\
	&		&		&		&		&		&		&		&		&		\\
5	&	92	&	1.8857 	&	1.8897 	&	1.8926 	&	1.8929 	&	0.0157 	&	1.9086 	&		&		\\
	&		&		&		&		&		&		&		&		&		\\
1	&	83	&	1.8375 	&	1.8423 	&	1.8450 	&	1.8454 	&	0.0140 	&	1.8593 	&		&		\\

\hline

\end{tabular}
\begin{tablenotes}
\item[] $^{\rm a}$ From T. Lennartsson by EBIT \cite{Lennartsson2013}
\item[] $^{\rm b}$ From T. Lennartsson by collisional-radiative model \cite{Lennartsson2013}
\item[] $^{\rm c}$ From Dipti et al by MCDF method \cite{doi:10.1139/cjp-2014-0636}
\end{tablenotes}
\end{threeparttable}
\end{table*}

In the Table ~\ref{Tab3}, the wavelength $\lambda$ (in nm) of E1 transitions between [Ne]3s$^{2}$3p$^{5}$3d$^{3}$ and [Ne]3s$^{2}$3p$^{6}$3d$^{2}$ configurations of W$^{54+}$ ion were presented with different active space. There are 110 levels of the excited states [Ne]3s$^{2}$3p$^{5}$3d$^{3}$. The full energy level list for both ground and the first excited state of W$^{54+}$ ion could be found in Table 1 of Ref. \cite{DingXB2016}. The \lq \lq index\rq \rq  \ in the table corresponds to the full index of the energy level. The results calculated from different active spaces were presented in column 3-6. The notation B+Q(Contri.) stands for the contribution from the Breit interaction and QED effects on the transition wavelength. The calculated results with Breit and QED effects were represented by C+B+Q. The other available experimental and theoretical results were also listed for comparison. It can be seen from Table ~\ref{Tab3} that the calculated transition wavelengths becomes converged with the increase of the active space. The Breit interaction and QED effects have important contribution (about 1\%) on the E1 transition wavelengths. The present results were in excellent agreement with the experimental data except for the first transition. According to the experiment, this observed line might affected by blending with other Ti-like tungsten transition and this explains the significant difference between the calculated wavelength and the experimental measurement. Comparing with the FAC results from T. Lennartsson et al. \cite{Lennartsson2013} and MCDF results by Dipti et al\cite{doi:10.1139/cjp-2014-0636}, the present calculation values were in reasonable agreement with their results. This agreement also indicate that the correlation model could be used for the first excited sates.

\begin{table*}

\renewcommand\arraystretch{1.1}
\footnotesize
\centering
\caption{The transition probabilities A$_{ji}$(in s$^{-1}$) in Babushikin gauge of some selected E1 transitions between [Ne]3s$^{2}$3p$^{5}$3d$^{3}$ and [Ne]3s$^{2}$3p$^{6}$3d$^{2}$ configurations in W$^{54+}$ ion. The i and j in \lq \lq Index \rq\rq is the energy level index which could be found in the Table 1 of \cite{DingXB2016}. The ratio of the transition probability calculated from Babushkin and Coulomb gauge were also presented.}\label{Tab4}
\begin{threeparttable}
\begin{tabular}{l c | c c c c c c}
\hline

 \multirow{2}*{i}	&	 \multirow{2}*{j}	&	 \multicolumn{6}{c}{A$_{ji}$(s$^{-1}$)  (C+B+Q)} 		\\
 \cline{3-8}
 \cline{3-8}
	&		&	DF	&	Ratio (DF)	&	n=3	&	Ratio (n=3)	&	n=5(5s-5d))	&	Ratio (n=5(5s-5d))	\\
\hline
1	&	30	&	8.495E+10	&	0.76 	&	7.836E+10	&	0.92 	&	8.500E+10	&	1.08 	\\
1	&	31	&	8.559E+11	&	0.66 	&	7.530E+11	&	0.86 	&	7.443E+11	&	1.09 	\\
1	&	32	&	3.203E+11	&	0.87 	&	4.776E+11	&	1.21 	&	5.911E+11	&	1.03 	\\
1	&	33	&	1.117E+12	&	0.74 	&	9.910E+11	&	0.99 	&	9.425E+11	&	1.06 	\\
1	&	34	&	1.061E+12	&	0.84 	&	6.954E+11	&	1.22 	&	5.184E+11	&	1.03 	\\
2	&	38	&	9.797E+11	&	0.81 	&	9.564E+11	&	1.02 	&	9.331E+11	&	1.05 	\\
1	&	38	&	2.826E+11	&	0.78 	&	3.272E+11	&	0.74 	&	2.958E+11	&	1.09 	\\
6	&	68	&	1.288E+12	&	0.78 	&	1.227E+12	&	0.93 	&	1.157E+12	&	1.07 	\\
7	&	79	&	1.282E+12	&	0.68 	&	1.218E+12	&	0.78 	&	1.167E+12	&	1.10 	\\
5	&	69	&	1.247E+12	&	0.68 	&	1.221E+12	&	0.74 	&	1.162E+12	&	1.10 	\\
8	&	110	&	4.825E+12	&	0.78 	&	4.155E+12	&	0.97 	&	4.031E+12	&	1.06 	\\
5	&	92	&	5.471E+12	&	0.80 	&	4.678E+12	&	1.03 	&	4.522E+12	&	1.05 	\\
1	&	83	&	5.820E+12	&	0.77 	&	5.256E+12	&	0.91 	&	5.089E+12	&	1.06 	\\

\hline

\end{tabular}
\begin{tablenotes}
\item[]
\end{tablenotes}
\end{threeparttable}
\end{table*}

In the Table ~\ref{Tab4}, the transition probabilities A$_{ji}$(in s$^{-1}$) for E1 transitions between [Ne]3s$^{2}$3p$^{5}$3d$^{3}$ and [Ne]3s$^{2}$3p$^{6}$3d$^{2}$ configurations in W$^{54+}$ ion with different active space in the Babushkin gauge and the ratio of the transition probability calculated from Babushkin and Coulomb gauge which corresponding to the length and velocity gauge in non-relativistic limit were presented. The agreement between the two results calculated in Babushkin and Coulomb gauge could indicate the accuracy of the calculated results to some extent. It can be found in the table that the ratio becomes unitary with the increase of the active space. It also indicate the wavefunction used in the present calculation was good enough and the most important correlation effects, Breit and QED effects were included in the present calculation.

\begin{figure}
\resizebox{0.5\textwidth}{!}{%
  \includegraphics{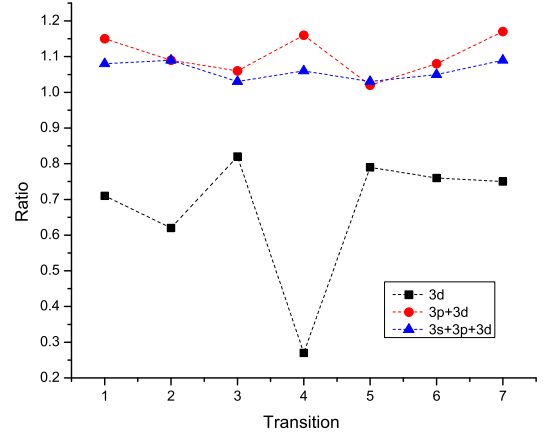}
}
\caption{The ratio of the transition probability of the first 7 transitions in Table \ref{Tab3} which calculated from Babushkin and Coulomb gauge in different correlation model. The black square, red circle and blue triangle were calculated with correlation from different subshells.}
\label{Fig:1}       
\end{figure}

Figure~\ref{Fig:1} shows the ratio of the transition probability of the first 7 transitions from Table ~\ref{Tab3} which calculated from Babushkin and Coulomb gauge by considering different correlation contribution from 3d, 3p and 3s subshells with the active space n=5(5s-5d). The ratio is far away from unitary when only 3d VV correlation was included, while it was improved much by considering 3d and 3p VV+CV+CC correlation. Finally, the ratio went around the unitary when the VV+CV+CC correlation from 3d, 3p and 3s subshells were all included. Therefore, the correlation from 3s and 3p were important for the transition probability.

%

\section{Conclusion}
\label{sec:3}
The energy levels, wavelengthes and  transition probabilities of the ground state and excited state of W$^{54+}$ ion were calculated by using multi-configuration Dirac-Fock method. A restricted active space was employed to investigate the electron correlation effects. The Breit interaction and QED effects were taken into account in relativistic configuration interaction calculation. The convergence of the energy levels and the transition wavelengthes were obtained with the increase of active space while the full  VV+CV+CC correlation was considered. The agreement of the transitiion probability calculated by different gauge indicate that the electron correlation effects were considered appropriately in the present calculation. Finally, the results shows that the correlation effects from {3s, 3p} orbitals have an important contribution to the energy levels, transition wavelength and probabilities for the ground state and excited state of W$^{54+}$ ion.

\section*{Acknowledgement}
This work was supported by National Nature Science Foundation of China, Grant No:11264035 and Specialized Research Fund for the Doctoral Program of Higher Education(SRFDP), Grant No: 20126203120004, International Scientific and Technological Cooperative Project of Gansu Province of China (Grant No. 1104WCGA186), the Young Teachers Scientific Research Ability Promotion Plan of Northwest Normal University (Grant No:NWNU-LKQN-15-3), JSPS-NRF-NSFC A3 Foresight Program in the field of Plasma Physics (NSFC: No.11261140328, NRF: 2012K2 A2A6000443).
%

%
\bibliographystyle{epj}
\bibliography{me}
%


%


\end{document}